\documentclass[aps,pre,amsmath,amssymb,floatfix,twocolumn,showpacs,superscriptaddress]{revtex4-1}
\usepackage{amssymb}
\usepackage{amsmath}
\usepackage[dvips]{graphicx}
\usepackage{mathrsfs}
\usepackage{ulem}
\usepackage[mathscr]{euscript}
\usepackage{gensymb}
\usepackage{color}
\usepackage{times}
\usepackage{dcolumn}
\usepackage{cases}
\usepackage{txfonts}
\usepackage{textcomp}
\usepackage[mathscr]{euscript}
\usepackage{indentfirst}
\usepackage{subfigure}
\usepackage{epstopdf}
\begin{document}
\title{Spin-ordered ground state and thermodynamic behaviors of the spin-$\frac{3}{2}$ kagome Heisenberg antiferromagnet}

\author{Tao Liu}
\affiliation{Theoretical Condensed Matter Physics and Computational Materials Physics Laboratory, School of Physics, University of Chinese Academy of Sciences, P. O. Box 4588, Beijing 100049, China}

\author{Wei Li}
\email{w.li@buaa.edu.cn}
\affiliation{Department of Physics, Key Laboratory of Micro-Nano Measurement-Manipulation and Physics (Ministry of Education), Beihang University, Beijing 100191, China}
\affiliation{International Research Institute of Multidisciplinary Science, Beihang University, Beijing 100191, China}

\author{Gang Su}
\email{gsu@ucas.ac.cn}
\affiliation{Theoretical Condensed Matter Physics and Computational Materials Physics Laboratory, School of Physics, University of Chinese Academy of Sciences, P. O. Box 4588, Beijing 100049, China}

\begin{abstract}
Three different tensor network (TN) optimization algorithms are employed to accurately determine the ground state and thermodynamic properties of the spin-3/2 kagome Heisenberg antiferromagnet. We found that the $\sqrt{3} \times \sqrt{3}$ state (i.e., the state with 120$^\circ$ spin configuration within a unit cell containing 9 sites) is the ground state of this system, and such an ordered state is melted at any finite temperature, thereby clarifying the existing experimental controversies. Three magnetization plateaus ($m/m_s=1/3, 23/27$, and $25/27$) were obtained, where the 1/3-magnetization plateau has been observed experimentally. The absence of a zero-magnetization plateau indicates a gapless spin excitation that is further supported by the thermodynamic asymptotic behaviors of the susceptibility and specific heat. At low temperatures, the specific heat is shown to exhibit a $T^2$ behavior, and the susceptibility approaches a finite constant as $T\rightarrow 0$. Our TN results of thermodynamic properties are compared with those from high temperature series expansion. In addition, we disclose a quantum phase transition between $q = 0$ state (i.e. the state with 120$^\circ$ spin configuration within a unit cell containing three sites) and $\sqrt{3}\times\sqrt{3}$ state in a spin-3/2 kagome XXZ model at the critical point $\Delta_c = 0.54$. This study provides reliable and useful information for further explorations on high spin kagome physics.

\end{abstract}
\maketitle

\section{Introduction}
Seeking quantum phases is always of great interest in condensed matter physics. It is widely thought that exotic quantum phases may appear, among others, in antiferromagnetic spin systems with frustration. The kagome Heisenberg antiferromagnetic (KHAF) model is one of the most frustrated antiferromagnets, which has therefore attracted extensive attention both theoretically and experimentally, increasing active debate in recent years. Earlier studies on the spin-1/2 quantum KHAF can be traced back to more than 20 years ago \cite{V. Elser1, J. B. Marston2, S. Sachdev3}. Through continuous efforts of many scientists, its ground state is generally believed to be a spin liquid without any symmetry breaking \cite{S. Yan4, S. Depenbrock5}. However, whether or not there is a gap in this interesting system still remains controversial \cite{S. Depenbrock5, T.H.Han6,Y.Iqbal7, Z. Y. Xie8, I. Rousochatzakis9, T.Liu10, W. Zhu11,T.H.Han12}. Meanwhile, the high spin ($S>1/2$) kagome physics has also gained great interest currently. For the spin-1 KHAF, the ground state was shown to be a nonmagnetic simplex valence bond crystal with geometric inversion symmetry breaking \cite{T. Liu13, W. Li14,H. J. Changlani15, T. Picot16, W. Li17, D. Ixert18}. For the spin-S KHAF, the magnetization curves \cite{D. Ixert19} up to $S = 2$ have been obtained with tensor network methods based on the infinite projected entangled pair states (iPEPS) \cite{F. Verstraete29, J. Jordan18}. In the limit of large $S$, the ground state of KHAF was argued to be the $\sqrt{3}\times\sqrt {3}$ state [that is the state with 120$^\circ$ spin configuration within a unit cell of area $\sqrt{3} a\times\sqrt {3}a$ containing nine sites as illustrated in Fig. 1(a)] with a long-range magnetic order \cite{O. Gotze19, A. L. Chernyshev20, A. L. Chernyshev25}.

In high spin kagome antiferromagnets,  the spin-3/2 KHAF is particularly intriguing. Intuitively, it should differ from those of spin-1/2 and spin-1 counterparts, and is also thought to be with the intermediate between quantum and classical nature. Nonetheless, despite a lot of efforts made both experimentally and theoretically, its nature is still ambiguous and controversial up to date. The experimental studies on a number of spin-3/2 KHAF materials such as $SrCr_{8}Ga_{4}O_{19}$ \cite{C. Mondelli21}, $SrCr_{8}Ga_{4-x}M_xO_{19}(M=Zn, Mg, Cu)$ \cite{S. E. Dutton22}, $Ba_{2}Sn_{2}ZnGa_{10-7p}Cr_{7p}O_{22}$ \cite{D. Bono23}, and Cr-jarosite $KCr_3(OH)_6(SO_4)_2$ \cite{Cr-jarosite24}, etc., have produced diverse results, leading to different even contradictory conclusions on the nature of the spin-3/2 KHAF. For instance, there are studies showing that it has no antiferromagnetic long-range order but undergoes a spin-glass transition \cite{C. Mondelli21,S. E. Dutton22,D. Bono23}, while some others reported that it possesses a long-range order with a nearly $120^\circ$ structure \cite{Cr-jarosite24,lee}. On the theoretical aspect, a direct study on overall ground-state as well as thermodynamic properties of the spin-3/2 KHAF is still sparse. The nonlinear spin-wave theory (NSWT) and real space perturbation theory \cite{A. L. Chernyshev25} gave the phase diagram of spin-S kagome XXZ model, revealing when the anisotropic parameter $\Delta$ increases, there is a phase transition from $q=0$ state \cite{A. B. Harris,A. Chubukov} [that is the state with $120^\circ$ spin configuration within a unit cell of area $a \times a$ containing three sites as depicted fin Fig. 1 (b)] to $\sqrt{3}\times\sqrt {3}$ state [Fig. 1(c)], where the critical point $\Delta_c$ is above 0.7. The coupled-cluster method \cite{O. Gotze26} showed that for the spin-3/2 kagome XXZ model, the transition from $q=0$ to $\sqrt{3}\times\sqrt {3}$ state happens at $\Delta_c=0.525$. The series expansion \cite{J. Oitmaa27} gives that the phase transition point is slightly lower than 0.8, which is close to the result obtained with the NSWT, but different from that obtained by the coupled-cluster method. In addition, it is still a challenge to accurately calculate the thermodynamic properties at low temperatures for kagome Heisenberg spin systems. The conventional high-temperature series expansion (HTSE) method \cite{A. Lohmann28} can effectively capture the nature at high or even intermediate temperatures, but was unable to reliably determine the thermodynamic behavior at low temperature. In this regard, a systematic and accurate study on the spin-3/2 KHAF is quite indispensable.

In this article, by employing the tensor network (TN)-based renormalization group method with three different optimization schemes, we determine with high accuracy the ground state and thermodynamic properties of the spin-3/2 quantum KHAF. We identify the $\sqrt{3} \times \sqrt{3}$ state, rather than the $q = 0$ state, is the ground state of this model, which is disordered at any finite temperature. We find that the magnetic curve exhibits $1/3$, $23/27$ and $25/27$ magnetization plateaus, where $1/3$-magnetization plateau has been observed in Cr-jarosite, but does not have a zero-magnetization plateau, indicating a gapless spin excitation. This is further supported by thermodynamic calculations, namely, the specific heat is shown to exhibit a dominant $T^2$ behavior at low temperatures, and the susceptibility approaches a finite constant as $T\rightarrow 0$. Extending the present KHAF model to the anisotropic XXZ case, we observed a quantum phase transition between $q=0$ and $\sqrt{3}\times\sqrt{3}$ states at the critical point $\Delta_c = 0.54$.

\begin{figure}[tbp]
  \includegraphics[angle=0,width=0.95\linewidth]{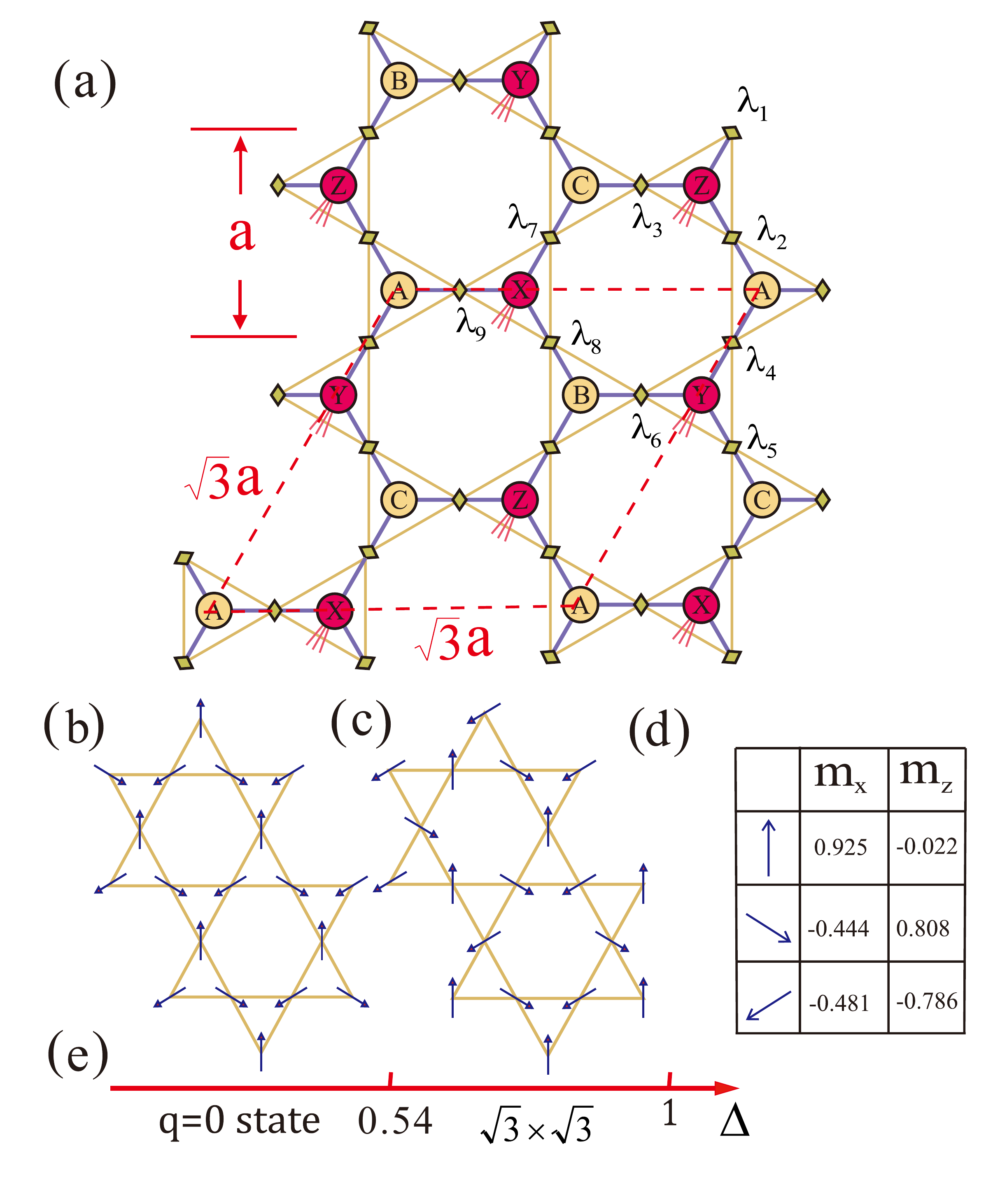}
  \caption{(Color online) (a) Tensor network representation of the spin-3/2 quantum KHAF. There are six inequivalent tensors, where each of tensors X, Y, and Z has three physical indices and three geometrical indices, and each of tensors A, B, and C has only three geometrical indices. Each bond has a diagonal matrix $\lambda_{i}$ $(i=1,2,...9)$, and the length unit $a=1$ is also indicated. The spin configurations of (b) $q=0$ and (c) $\sqrt{3}\times\sqrt{3}$ states are depicted. (d) $m_{x}$ and $m_{z}$ components of the local magnetization obtained by cluster update ($D=16$); the three orientations are 120$^{\circ}$ to each other. We schematically plot them as pointing up, right-down, and left-down, respectively. (e) A phase diagram for the spin-3/2 quantum XXZ model with the anisotropic parameter $\Delta$ on kagome lattice. When $0\leq\Delta < 0.54$, the ground state should be in the $q=0$ state; when $0.54\leq \Delta \leq 1$, the ground state is in the $\sqrt{3}\times\sqrt{3}$ state. A quantum phase transition occurs at the critical point $\Delta_{c}= 0.54$.}
  \label{fig1}
\end{figure}

\section{Tensor Network Simulations}

\subsection{Model and methods}
Consider a quantum spin-3/2 KHAF, whose Hamiltonian reads
\begin{equation}
H = J \sum\limits_{<i,j>}\textbf{S}_{i}\cdot\textbf{S}_{j}-h\sum\limits_{i}\textbf{S}^{z}_{i},
\label{hamil}
\end{equation}
where $J$ is the antiferromagnetic coupling constant, $\textbf{S}_{i}$ stands for the spin operator with S=3/2 on \textit{i}-th site, and $\langle i,j \rangle$ denotes the summation over nearest neighbors. The ground state $|\psi_g\rangle$ of Eq. (\ref{hamil}) can be written in the form of iPEPS for a variational study.

The imaginary-time evolution, via Trotter-Suzuki decomposition, TSD \cite{Masuo Suzuki30}, can be used to get the optimized $|\psi_g\rangle$, i.e., $|\psi_{g}\rangle=\lim_{\beta\rightarrow \infty} e^{-\beta H}|\psi_\rangle$, where $|{\psi}\rangle$ is the starting random state. Following TSD, $ e^{-\beta H}=[\prod_{a,b}e^{-\tau h_{a}}e^{-\tau h_{b}}]^{K}$, where $K\tau = \beta$, and $h_{a}$ and $h_{b}$ represent the local Hamiltonian of the upper and lower triangles, respectively. In calculating the ground state properties, we first set the Trotter step $\tau=0.1$  and reduce it gradually to $10^{-5}$. In calculating the  thermodynamic properties (the details of algorithm are included in the Appendix), we set $\tau=0.01$ for convenience, and take the second-order TSD to further reduce the Trotter error.

Figure 1(a) presents the iPEPS representation of the wave function, which contains six inequivalent tensors X, Y, Z and A, B, C.
%Each of X, Y and Z tensors contains three geometrical indices and three physical indices. A diagonal matrix $\lambda_{i}$, $i=1,2,...9$ (denoted by diamond) resides on each geometrical bond of the TN.
When one applies $e^{-\tau h_{a(b)}}$ to the tensor-network trial wavefunction, the geometric bond dimension will be increased continually. To truncate the bond, restricting its dimension as $D$, the influence of environment on the bond should be considered. As the exact environment is generally not feasible, proper approximation schemes are needed, which include local optimization (simple update \cite{H. C. Jiang31} and cluster update \cite{W. Li32, L. Wang33}) and global optimization (full update \cite{J. Jordan18,H. N. Phien35,M. Lubasch36,ran38,R. Orus32}), which are all adopted in our calculations.

\subsection{Ground state} 
\begin{figure}
\includegraphics[width=0.95\linewidth,clip]{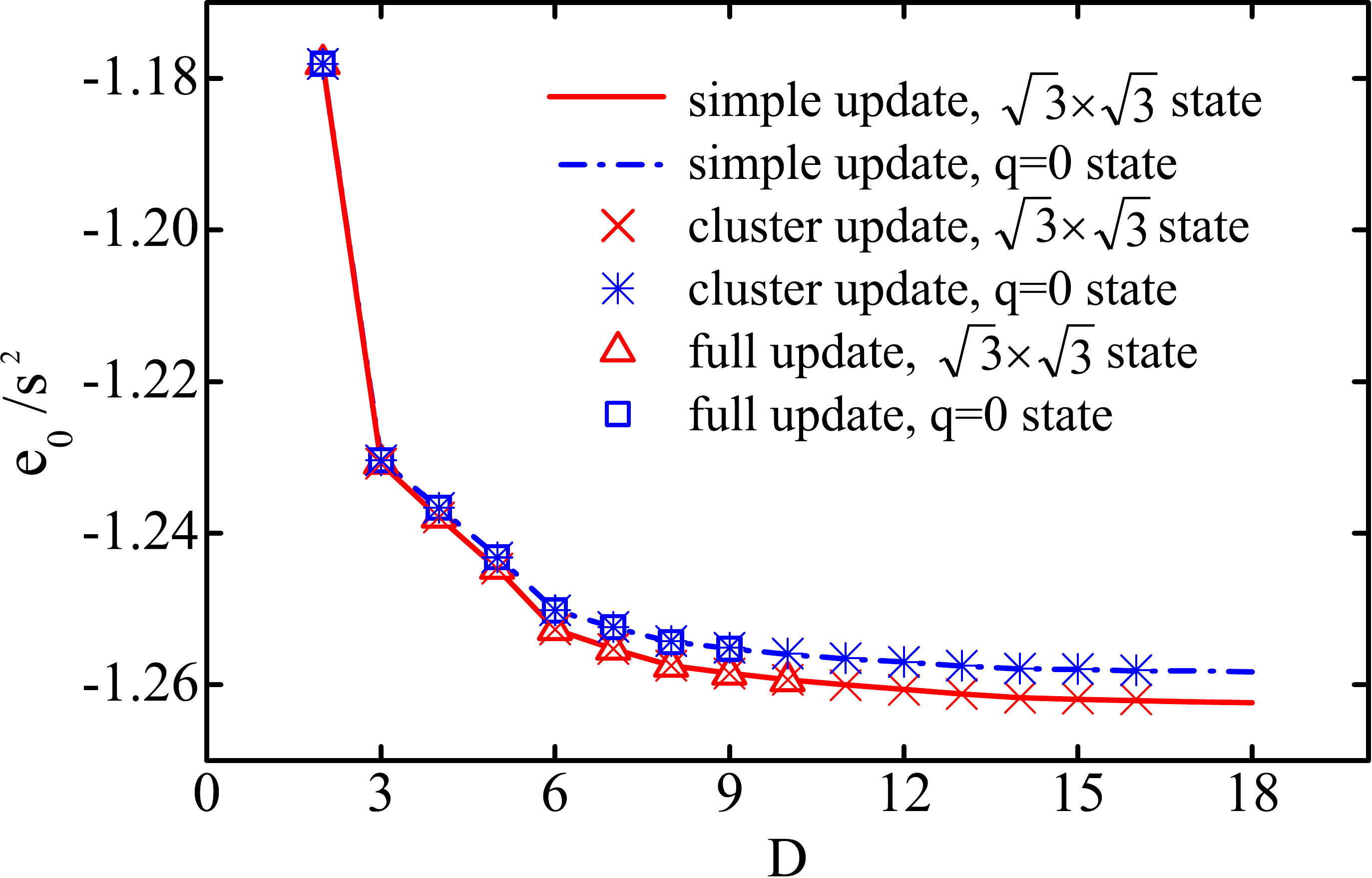}
\caption{(Color online) The energies of the $\sqrt{3}\times\sqrt{3}$ and $q=0$ states as function of bond dimension $D$ calculated by three different (simple, cluster, and full) update schemes for the spin-3/2 KHAF model.}
\label{fig2}
\end{figure}

Considering the two candidate spin ordered states illustrated in Fig. \ref{fig1} (a), i.e., the $\sqrt{3}\times\sqrt{3}$ state with a unit cell of size $\sqrt{3}a\times \sqrt{3}a$, and the $q=0$ state with a unit cell of size $a\times a$, we adopted the $\sqrt{3}\times \sqrt{3}$ tensor network structure which can represent both states. In practical calculations with relatively small $D$, we found that, if one starts with a random initial wave function $|\psi_{0}\rangle$ and takes a variational optimization $|\psi_{g}\rangle= \lim_{\beta \rightarrow \infty}e^{-\beta H}|\psi_{0}\rangle$, one will have the chance to get both $q=0$ and $\sqrt{3}\times\sqrt{3}$ states, depending on the initial wave functions. This implies that the two states are very competitive, but are not degenerate, for our calculations show that in large D calculations the $\sqrt{3}\times\sqrt{3}$ state turns out to bear an energy clearly lower than that of the $q=0$ state (Fig. \ref{fig2}). In order to make the results more reliable, we also tested the tensor network structure with different unit cells by the simple update scheme, and did not find any state with energy lower than the $\sqrt{3}\times\sqrt{3}$ state.

In Fig. 2, we present the results of energy using the iPEPS algorithm with three different optimization schemes (simple, cluster and full update), which give accordant results, showing that the ground state is the $\sqrt{3}\times\sqrt {3}$ state. The ground state energy $e_{0}/s^{2}=-1.26237$ for bond dimension $D=18$. By extrapolating the result to $D=\infty$, we get the ground state energy estimated as low as $-1.265(2)$, which is very close to the extrapolated value $-1.2680$ obtained by the coupled cluster method \cite{O. Gotze26}.

\subsection{Magnetization} To see how the spin configurations of the spin-3/2 KHAF are arranged in the ground state, we calculated the local magnetization and found a $\sqrt{3}\times\sqrt{3}$ magnetic structure, which is depicted  in Fig. \ref{fig1} (c). There are three different kinds of local spin orientations, and the corresponding $x$ and $z$ components of local magnetization ($m_x$ and $m_z$) are listed in Fig. \ref{fig1} (d), $m_{y}=0$ due to the choice of real wave functions in the calculations. We also checked the calculations with complex wave functions, which gives the result of $\sqrt{3}\times\sqrt{3}$ order as well.

In Fig. 3, we show the ground-state magnetization curve obtained by full update with $D=5$, where three magnetization plateaus are found, i.e., $m/m_s=1/3$, $23/27$, and $25/27$, with $m_s$ the saturation magnetization. It is the consequence of the enlarged unit cell of $\sqrt{3}\times\sqrt{3}$ state on kagome lattice, consistent with the topological quantization condition of $n(S-m)=$integer with $n=9$. We note, interestingly, there is a jump from $25/27$ plateau to saturation, consistent with the observation of macroscopic magnetization jumps due to independent magnons \cite{schnack}. In Ref. [\onlinecite{D. Ixert19}], these three magnetization plateaus were also found with the iPEPS algorithm, suggesting that the present calculations and findings are reasonable, although we note that the widths of the plateaus are different, especially for the $23/27$-plateau. It is also interesting to point out that a 1/3-magnetization plateau-like anomaly was experimentally observed at $T$=1.3 K in $S$=3/2 perfect kagome lattice antiferromagnet KCr$_3$(OH)$_6$(SO$_4$)$_2$ by Okuta \textit{et al}. \cite{Cr-jarosite24}, which is compatible with our calculations. Besides, there is no $m=0$ plateau in the magnetic curve, indicating the existence of gapless Goldstone modes in spin excitations, which constitutes a strong indication of spontaneous continuous symmetry breaking in the ground state. However, such a $\sqrt{3}\times\sqrt{3}$ ordered state is found to be melted at finite temperatures, in agreement to the Mermin-Wagner theorem \cite{Mermin37}. The experimental observation of the antiferromagnetic order below the Ne\'el temperature T$_N$=4.5 K in Cr-jarosite by the magnetic susceptibility measurement by Okuta \textit{et al}. \cite{Cr-jarosite24} may be ascribed to a three-dimensional effect at low temperatures.

\begin{figure}
\includegraphics[width=0.95\linewidth,clip]{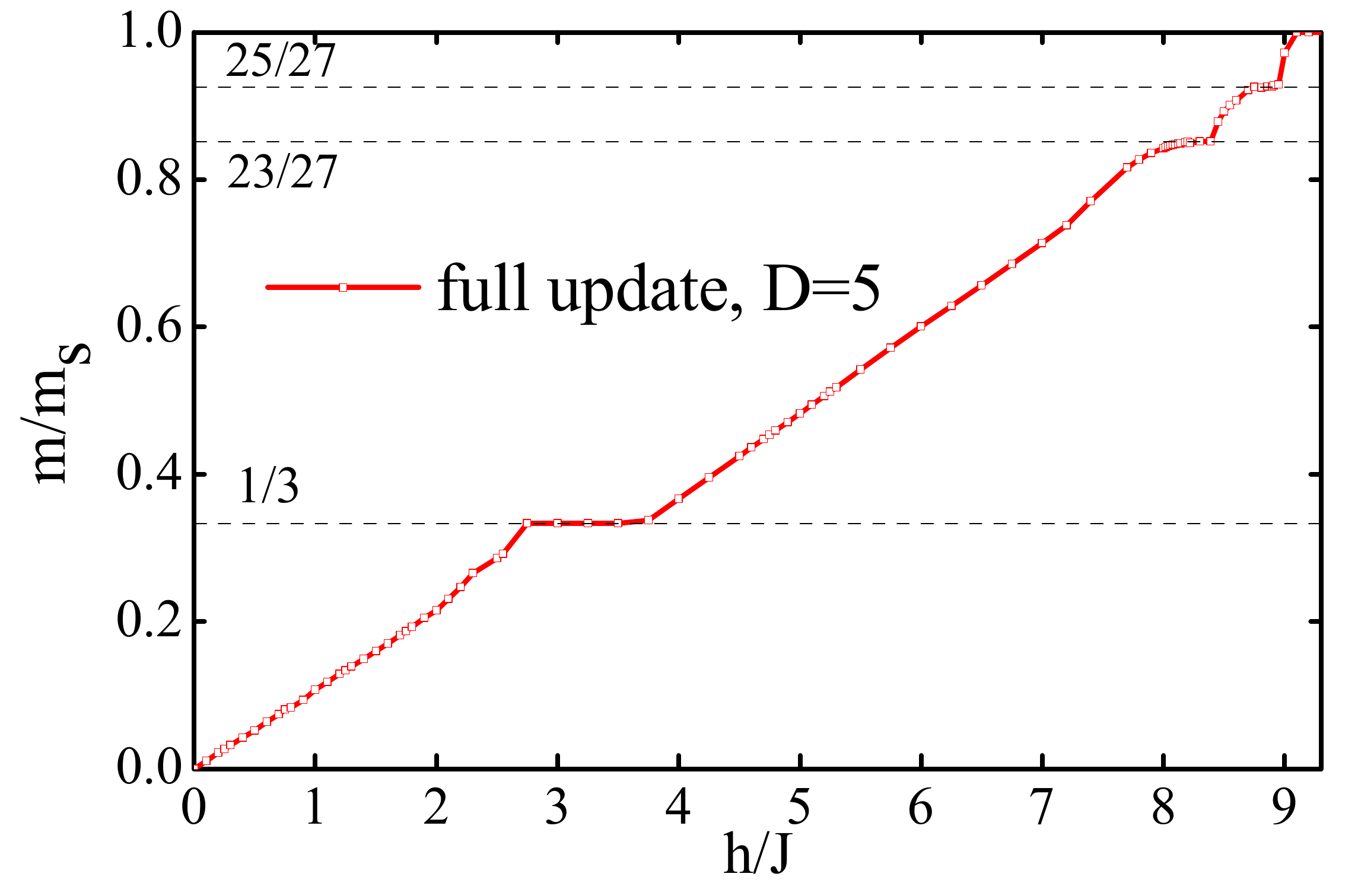}
\caption{(Color online) The magnetization curve $m/m_s$ ($m_s$ the saturation magnetization) versus the magnetic field, where there exist three magnetization plateaus, i.e., $m/m_s=1/3$, $23/27$, and $25/27$, but no $m=0$ plateau.}
\label{fig3}
\end{figure}

 \begin{figure}
\includegraphics[width=0.95\linewidth,clip]{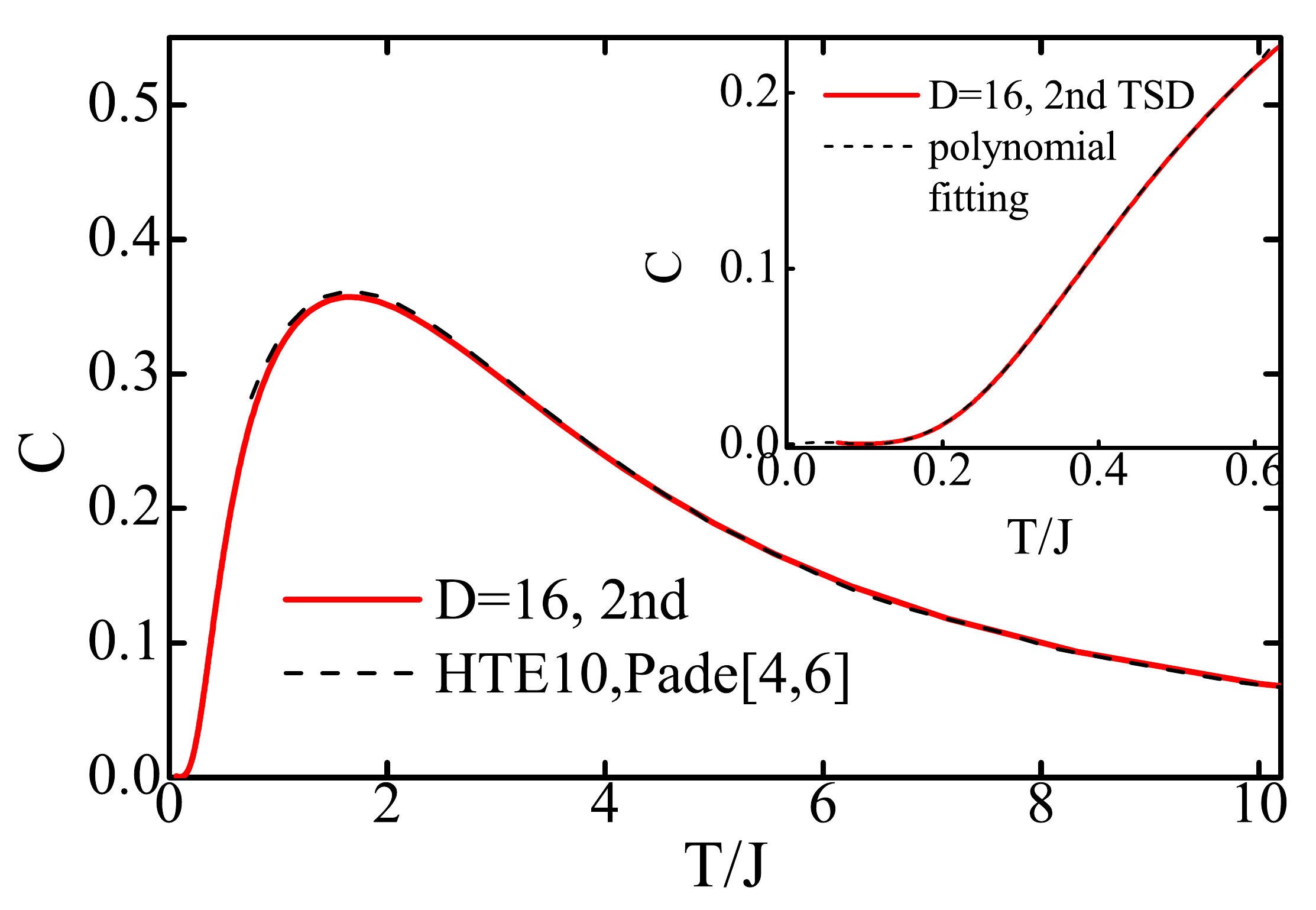}
\caption{(Color online) Specific heat $C$ versus $T$ for the spin-3/2 KHAF model. The solid (red) line represents the results obtained by the cluster update (second-order TSD, with $D=16$). The dashed (black) line is calculated by the {\it ten}-th order high temperature series expansion. Inset: the low-temperature part of $C$, which can be well fitted with a polynomial $C=7.80(T/J)^2-64.7(T/J)^{5/2}+188(T/J)^3-222(T/J)^{7/2}+92.4(T/J)^4$.}
\label{fig4}
\end{figure}

\subsection{Specific heat}  Figure 4 gives the specific heat $C$ as a function of temperature $T$ for the spin-3/2 KHAF, which was calculated by the thermal tensor-network algorithm \cite{ran38, Li-2011, Ran-2013} with cluster update scheme ($D=16$). For a comparison, we also include the results of ten-th order HTSE. It can be seen that the specific heat shows a broad peak at around $T/J\sim 2$, reflecting a typical feature of the 2D isotropic antiferromagnets. Our results are in agreement with those from HTSE at high temperatures, verifying the reliability of our method. At low temperatures, the specific heat shows an algebraic behavior, and the fitting result gives $C=7.80(T/J)^2-64.7(T/J)^{5/2}+188(T/J)^3-222(T/J)^{7/2}+92.4(T/J)^4$, as shown in the inset of Fig. 4. As $T\rightarrow 0$, the specific heat behaves in the form of $C(T) \sim T^2$, which also supports that the low-lying excitation is gapless.

 \begin{figure}
\includegraphics[width=0.95\linewidth,clip]{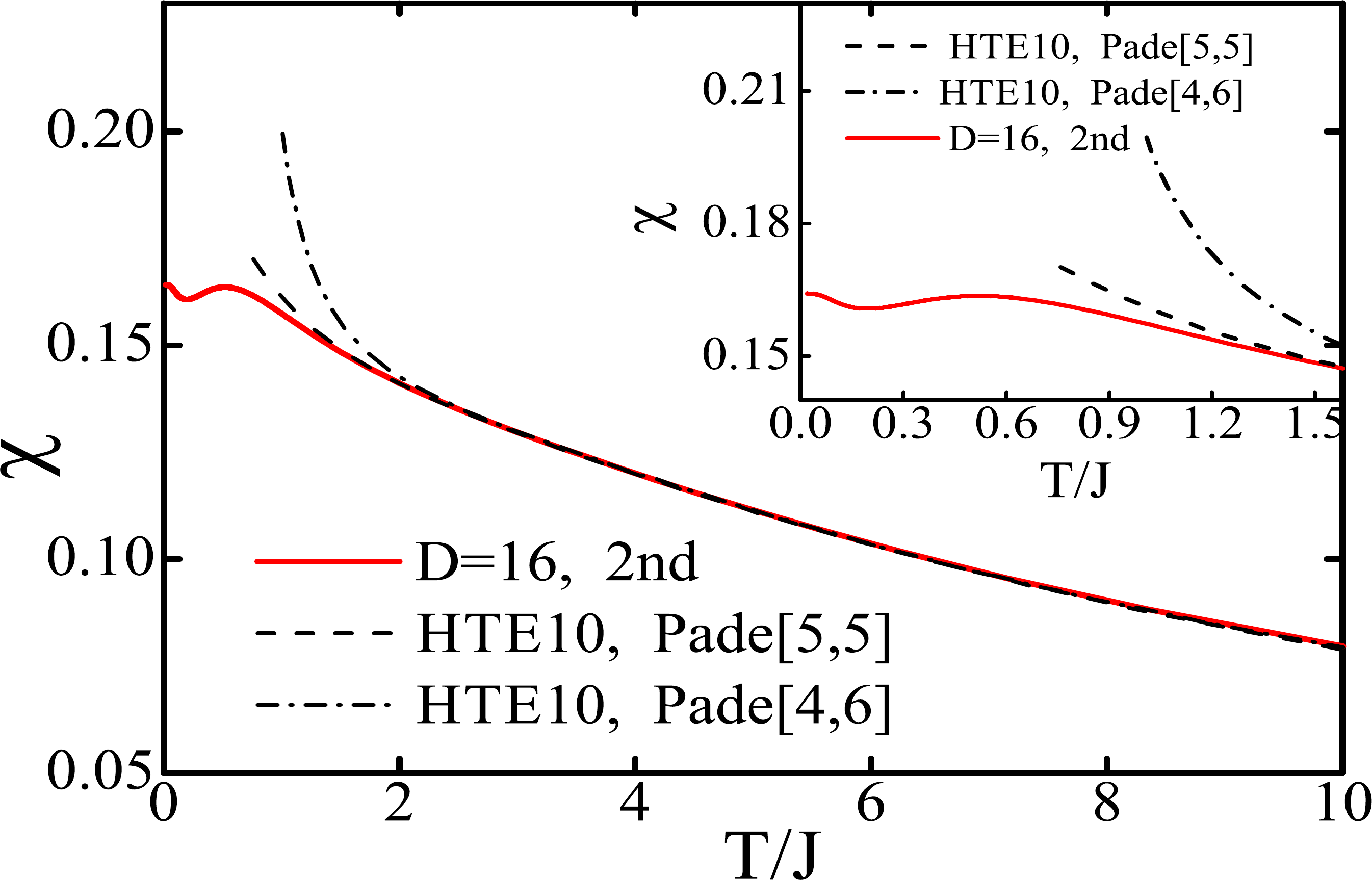}
\caption{(Color online) Zero-field magnetic susceptibility $\chi$ as a function of temperature $T$ for the spin-3/2 KHAF model. The result (red solid line) is obtained by the cluster update scheme (second-order TSD and $D=16$). The dashed and dot-dashed lines (black) are calculated by the ten-th order high temperature series expansion under different parameters Pade[5,5] and Pade[4,6]. Inset: the low-temperature part of $\chi$ versus $T$, showing that at $T\rightarrow 0$, $\chi\rightarrow 0.164$.}
\label{fig5}
\end{figure}

\subsection{Susceptibility} Figure 5 presents the susceptibility $\chi$ obtained with the cluster update scheme ($D=16$), it is observed that $\chi (T)$ exhibits an interesting behavior with a dip and a peak at low temperatures, and obeys a Curie-Weiss law at high temperatures. The low-temperature peak of $\chi (T)$ is the typical feature of an antiferromagnet. As $T\rightarrow 0$, $\chi (T)$ does not converge to zero but to a finite constant 0.164, as demonstrated in the inset of Fig. 5, due to the low-lying gapless magnetic excitations. We also include HTSE results \cite{A. Lohmann28} at two different parameters Pade[5,5] and Pade[4,6] as comparisons. Both results are consistent with our calculations at high temperatures, while they deviate remarkably from the iPEPS result at relatively low temperatures, revealing the failure of HTSE in that regime.

\subsection{Spin-3/2 kagome XXZ model} 
\begin{figure}
\includegraphics[width=0.98\linewidth,clip]{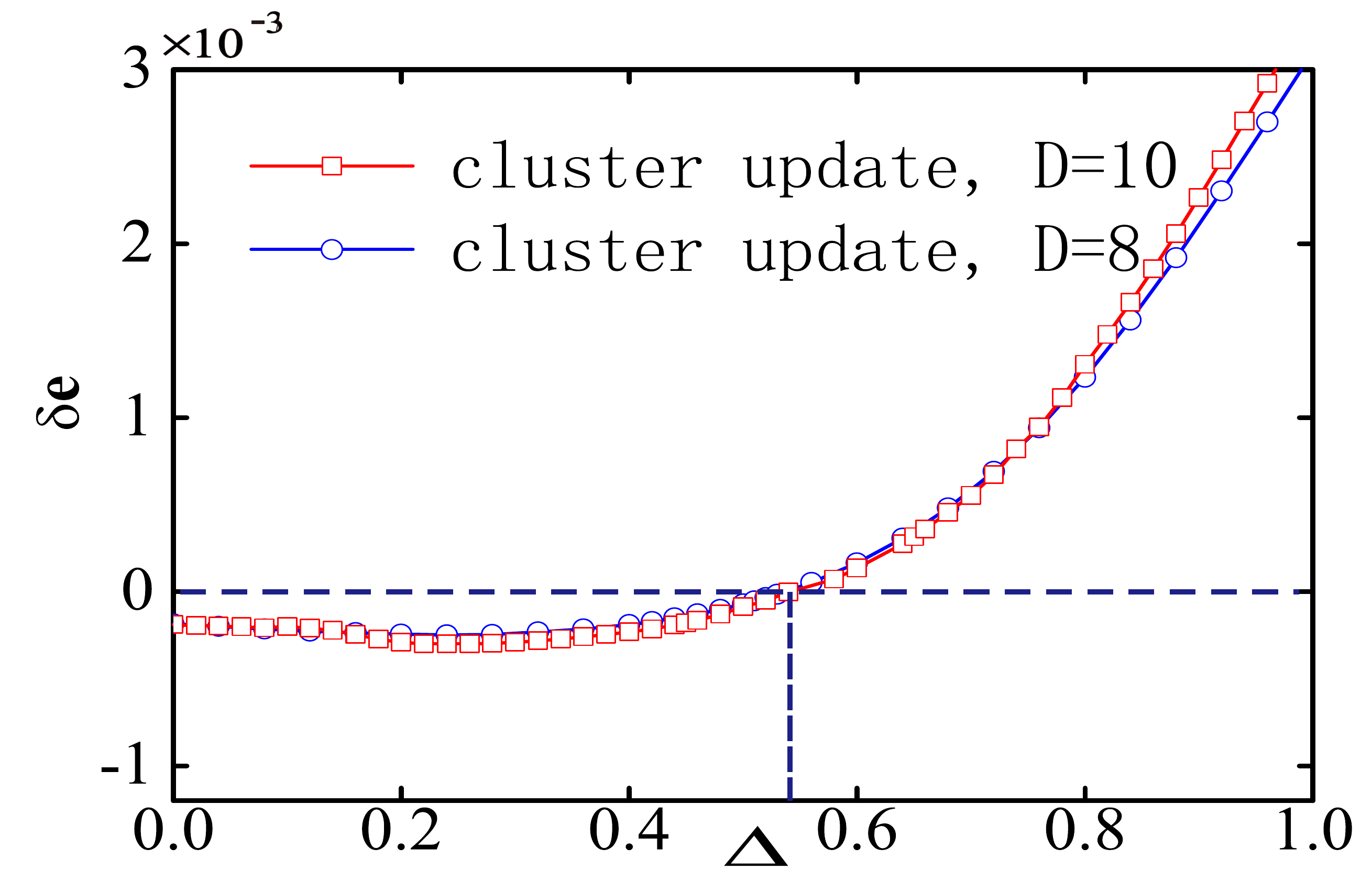}
\caption{(Color online)  The energy difference $\delta e=(e^{0}_{q=0}-e^{0}_{\sqrt{3}\times\sqrt{3}})/s^{2}$ between $q=0$ and $\sqrt{3}\times\sqrt{3}$ states as a function of $\Delta$. The phase transition point is $\delta=0.54$.}
\label{fig6}
\end{figure}

Now let us address the quantum phase transition of the spin-3/2 kagome XXZ model. The corresponding Hamiltonian is $H=\sum_{<i,j>}(S^{x}_{i}S^{x}_{j}+S^{y}_{i}S^{y}_{j}+\Delta S^{z}_{i}S^{z}_{j})$ with $0\leq\Delta\leq1$ the anisotropic parameter. We show, in Fig. \ref{fig6}, cluster update results ($D=10$) of the energy difference between the $q=0$ and $\sqrt{3} \times \sqrt {3}$ states as a function of $\Delta$. A quantum phase transition occurs at $\Delta_{c}= 0.54$: When $\Delta < 0.54$, the energy of $q=0$ state is lower than that of the $\sqrt{3} \times \sqrt {3}$ state, suggesting that the ground state is $q=0$ ordered; However, when $\Delta > 0.54$, the ground state is $\sqrt{3} \times \sqrt {3}$ ordered. A phase diagram for the spin-3/2 kagome XXZ model is depicted in Fig. \ref{fig1} (e). Note that the quantum critical point $\Delta_{c}= 0.54$ obtained by the iPEPS is lower than $\Delta_{c}= 0.72235$ determined by the NSWT \cite{A. L. Chernyshev25} and is more close to $\Delta_{c}= 0.525$ determined by the coupled cluster method  \cite{O. Gotze26}.

\section{Conclusion} A systematic tensor-network study on the ground-state and thermodynamic properties of the spin-3/2 KHAF is performed by using three optimization schemes. We identify that the $\sqrt{3} \times \sqrt{3}$ state is the ground state of this system, which is melted at finite temperatures. Three magnetization plateaus at $1/3$, $23/27$, and $25/27$ are found in the magnetic curve, where the $1/3$-magnetization plateau has been observed in the compound KCr$_3$(OH)$_6$(SO$_4$)$_2$. The absence of zero-magnetization plateau, the algebraic low-$T$ specific heat $C$, together with the fact that zero-field susceptibility $\chi (T)$ tends to a finite nonzero constant 0.164 as $T\rightarrow 0$, all suggest the existence of a gapless spin excitation. In addition, we observed a quantum phase transition between $q = 0$ and $\sqrt{3}\times\sqrt{3}$ states in the spin-3/2 kagome XXZ model at the critical point $\Delta_c = 0.54$. The present study not only provides useful insight into the high spin kagome physics but also resolves some ambiguities concerning the spin-3/2 kagome antiferromagnets.

\begin{acknowledgments}
We are indebted to S.-J. Ran, C. Peng, X. Chen, Q. B. Yan, Z. C. Wang, Q. R. Zheng, and Z. G. Zhu for stimulating discussions. This work was supported in part by the MOST of China (Grant  No. 2012CB932900 and No. 2013CB933401), the NSFC (Grant No. 14474279 and No. 11504014) and the Strategic Priority Research Program of the Chinese Academy of Sciences (Grant No. XDB07010100), WL was also supported by the Beijing Key Discipline Foundation of Condensed Matter Physics.
\end{acknowledgments}

\begin{appendix}
\setcounter{equation}{0}
\setcounter{figure}{0}
\setcounter{table}{0}
\makeatletter
\renewcommand{\theequation}{A\arabic{equation}}
\renewcommand{\thefigure}{A\arabic{figure}}

\section{Tensor Network Update Algorithms}
In the Appendix, we describe briefly the algorithmic details with
simple \cite{H. C. Jiang31}, cluster \cite{W. Li32, L. Wang33}, and full update \cite{J. Jordan18,H. N. Phien35,M. Lubasch36} schemes in optimizing the iPEPS wave function of
the spin-3/2 KHAF model.

\subsection{Simple update}

\begin{figure}
\includegraphics[width=0.95\linewidth,clip]{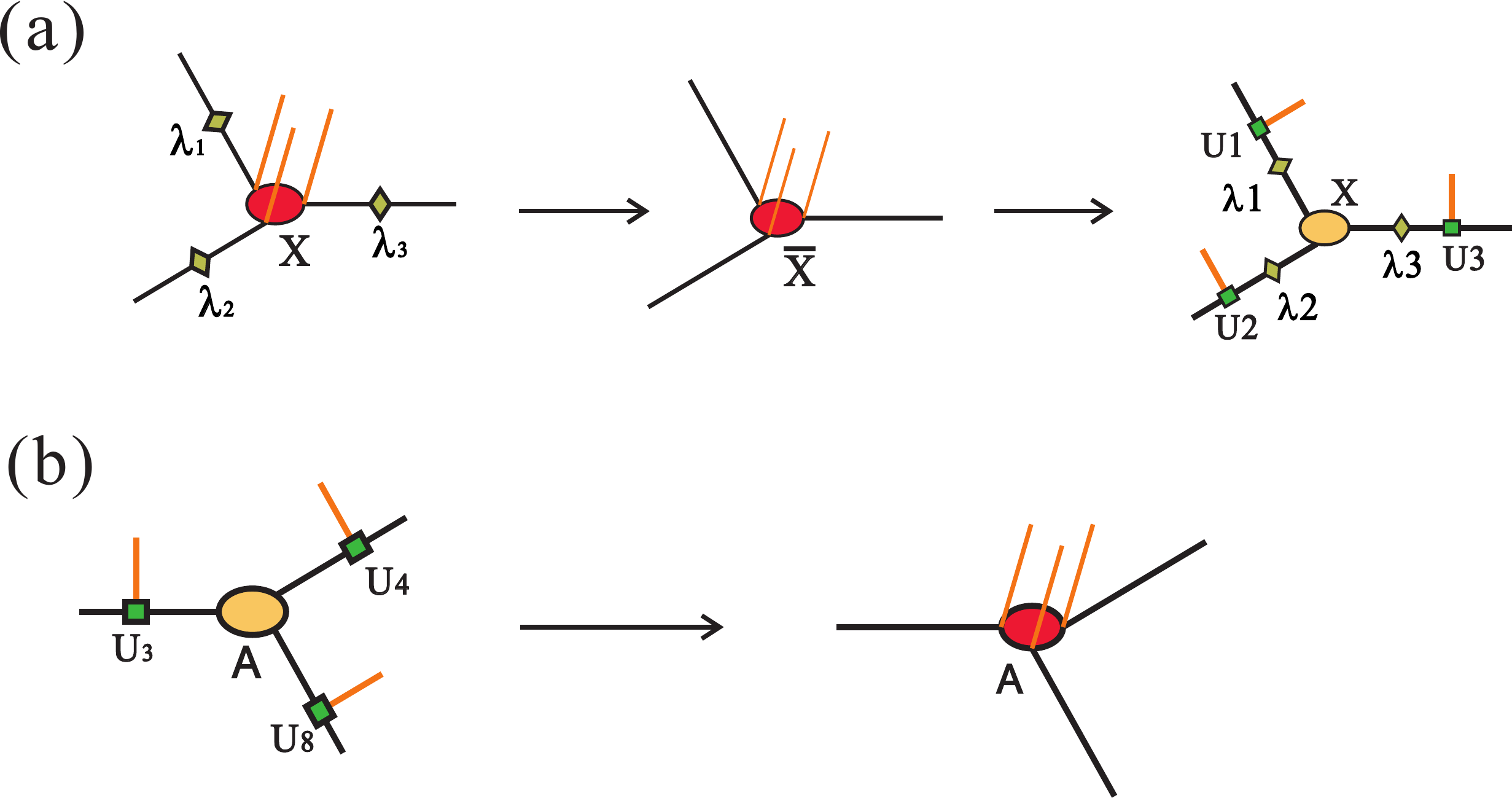}
\caption{(Color online) Graphic representation for the simple update scheme. (a)
The imaginary time evolution for tensor X; the tensors Y and Z are treated
similarly. (b) Operations on tensor A; tensors B and C are treated similarly. }
\label{simple}
\end{figure}

As shown in Fig. \ref{simple}(a), we take the tensor X as the ``system", three $\lambda$'s around X as the effective ``environment". We first absorb the $\lambda$'s into the
tensor X:

\begin{equation}\label{1}
\bar{X}^{p_{1}p_{2}p_{3}}_{i_{1}i_{2}i_{3}}=\sum_{j_{1}j_{2}j_{3}}X^{p_{1}p_{2}p_{3}}_{j_{1}j_{2}j_{3}}\lambda^{1}_{i_{1}j_{1}}...
\lambda^{2}_{i_{2}j_{2}}\lambda^{3}_{i_{3}j_{3}}.
\end{equation}

Then we perform the imaginary-time evolution. By acting the projection operator
$e^{-\beta h_{a(b)}}$ onto $\bar{X}$, we get the renewed $\bar{X}$. Next, we
decompose the physical index from the tensor $\bar{X}$  using higher-order
singular value decomposition
in the way of (taking $M^{1}$ as an example)

\begin{equation}\label{2}
  M^{1}_{p_{1}i_{1},p_{1}^{'}i_{1}^{'}}=\sum_{i_{2}i_{3},p_{2}p_{3}}\bar{X}^{p_{1}p_{2}p_{3}^{'}}_{i_{1}i_{2}i_{3}^{'}}
  ...
  \bar{X}^{p_{1}p_{2}p_{3}}_{i_{1}i_{2}i_{3}},
\end{equation}

\begin{equation}
  [U_{1},S_{1}]=\mathrm{eig}(M^{1}).
\label{3}
\end{equation}

The reduced density matrices $U_{i}$ $(i=1,2,3)$ and the diagonal matrices
$S_{i}$ can be obtained by the eigenvalue decomposition. In this process, we
keep $D$ largest diagonal elements of $S_{i}$ and corresponding $D$ columns in
$U_{i}$ to ensure that the computing cost does not continue to increase, and
then update $\lambda$'s with these $\sqrt{S}$ matrices.

Subsequently, we contract the tensor A with the three transform matrices $U$'s,
as shown in Fig. \ref{simple}(b). Likewise, the tensors Y, Z and B, C are also
be updated in this way. This process moves the physical indices
from tensors X, Y, Z to tensors A, B, C, which completes the projection substep
on the upper  triangle. Now we continue to perform similar operations again and
complete the projection of a down  triangle, moving the physical indices back to
the tensors X, Y, Z. This accomplishes
the projection of a full Trotter slice, while the structure of the iPEPS
representation remains intact. We set $\tau=0.1$ at the beginning and gradually
reduce it to $\tau=10^{-6}$ until the iPEPS converges.

\subsection{Cluster update}

\begin{figure}
\includegraphics[height=1 \linewidth,clip]{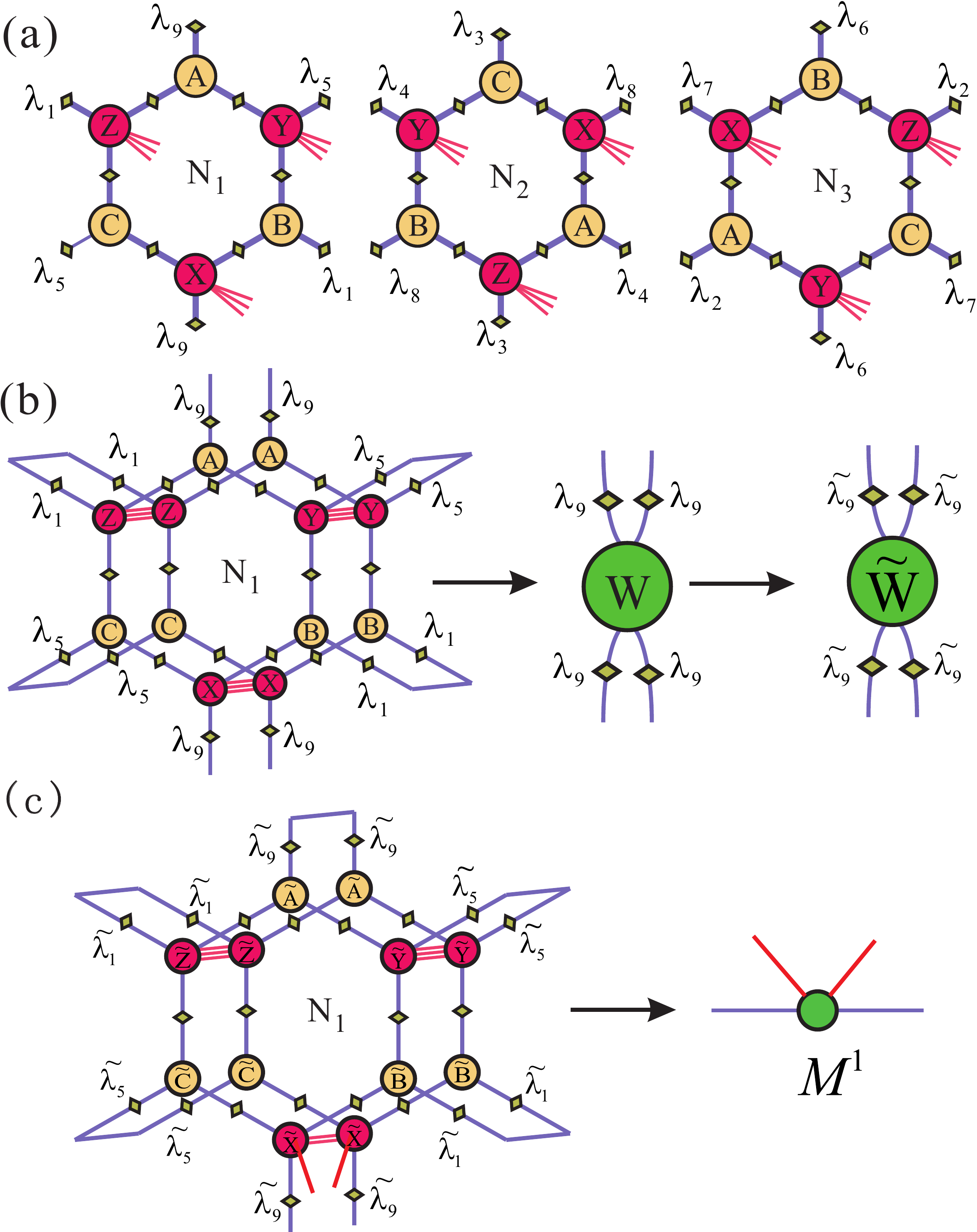}
\caption{(Color online) Graphical representation for the cluster update scheme.
(a) The tensor network contains three inequivalent kinds of hexagons, marked by
$N_{1}$, $N_{2}$, $N_{3}$, respectively. Each hexagon contains nine physical
indices and six geometrical indices. (b) Take unit cell $N_{1}$ as an example.
We contract the double-layer tensor cluster to get tensor W, which will later be
canonicalized for updating the effective environment $\tilde{\lambda_{9}}$, as well as the tensors $X$ and
  $A$. (c) After the ``canonicalization", we contract the hexagon tensor cluster
  with its conjugate layer, leaving only $\tilde{\lambda_{9}}$ and one physical
  bond open, and obtain the reduced density matrix
  $M^{1}$.}
\label{cluster}
\end{figure}

In the simple update scheme, only three $\lambda$ matrices around a single
tensor (A, B, C, X,Y, Z) are considered to approximately represent the effect of
environments. In the cluster update, we choose a hexagon that consists of six tensors
as a cluster. As show in Fig. \ref{cluster} (a), any hexagon of the tensor
network contains six adjacent tensors X, Y, Z, A, B, and C. One can see, after
some counting, that there are three kinds of hexagons $N_{1}$, $N_{2}$, and
$N_{3}$. Just like in the simple update, we consider the environment ``locally",
while the difference is that the system is extended from a single tensor X (Y or
Z) to a hexagon $N_{1}$ ($N_{2}$, or $N_{3}$). We first make the cluster tensors
to satisfy the
orthogonal conditions \cite{ran38} (taking the cluster tensor $N_{1}$ for
example) and then consider the six $\lambda$'s on the dangling bonds
as the ``environment" of the hexagon cluster for cutting process in the
following:

\begin{eqnarray}\label{4}
  \sum_{p}\sum_{g_{1}g_{2}\cdots g_{i-1}g_{i+1}\cdots
  g_{n}}N_{p,g_{1}g_{2}\cdots g_{i}\cdots g_{n}}N_{p,g_{1}g_{2}\cdots
  g^{'}_{i}\cdots g_{n}} \nonumber \\=\delta_{g_{i}g^{'}_{i}}\lambda^{2}_{g_{i}}.
\end{eqnarray}

Now we show how to make the cluster tensors $N_{1}$, $N_{2}$ and $N_{3}$ satisfy
the orthogonality conditions simultaneously. As shown in Fig. \ref{cluster} (b),
by taking $N_{1}$ as an example, we build a double-layer structure of the
cluster tensor and connect all geometric bonds on
bra and ket layers except a certain bond (e.g., $\lambda_{9}$). Then, we
contract the left graph of Fig. \ref{cluster} (b) to get a
matrix W [as in the middle graph of Fig. \ref{cluster} (b)]. For a later
convenience, we do not absorb the diagonal matrix $\lambda_{9}$ into W.
Subsequently, we suppose W and $\lambda_{9}$ comprise a one-dimensional chain
and perform a $1D$ canonicalization procedure\cite{R. Orus32}
to get a renewed $\tilde{\lambda}_{9}$ and $\tilde{W}$, and what we have done to
$W$ is actually acting on tensors X and A and we get
renewed tensors $\tilde{X}$ and $\tilde{A}$. In this way, we complete the
regularization on the cluster tensor $N_{1}$ along the direction of
$\lambda_{9}$ and obtain the renewed cluster tensor $N_{1}$ [Fig.
\ref{cluster}(c) left]. We then continue to regularize the cluster tensor
$N_{1}$ in the direction of $\lambda_{5}$ and get renewed tensors $\tilde{Y}$,
$\tilde{C}$, and $\tilde{\lambda_{5}}$. Then, we regularize the cluster tensor
$N_{1}$ in the direction of $\lambda_{1}$ and renew tensors $\tilde{Z}$,
$\tilde{B}$, and $\tilde{\lambda_{1}}$. We iterate this procedure until the
so-obtained renewed $N_{1}$ simultaneously satisfies the orthogonality
conditions in all three directions. Similar operations can apply to the cluster
tensors $N_{2}$ and $N_{3}$. In the end, such an arbitrary hexagon cluster
tensor satisfies the orthogonality condition. Note that we do not invoke any
truncation so far, and what we have done is just to regularize cluster tensor
$N_{1}$, $N_{2}$, $N_{3}$ into their ``canonical" form.

Next we must move the physical indices from tensors X, Y, Z to A, B, C. This
procedure will increase the bond dimensions, and one needs to truncate them
back. We again take the cluster tensor $N_{1}$ as an example. The six
$\tilde{\lambda}$'s around $N_{1}$ can be regarded as the environment. As shown
in Fig. \ref{cluster} (c), we contract $N_{1}$ with its conjugate and obtain the
reduced density matrix $M^{1}$. Then taking an eigenvalue decomposition on $M^{1}$
[see Eq.(\ref{3})], we get the isometry $U_{1}$ and diagonal matrix $S_{1}$. By
the same procedure we get $U_{2},...,U_{9}$, and $S_{2},...,S_{9}$. The
subsequent steps are in the same way as in the simple update scheme [see Eq. (\ref{3})
and the paragraph under it].

Finally, after we obtain the PEPS representation of the converged ground state
wave function, we need to measure the physical quantity. Here we adopt the
infinite time-evolution block decimation (iTEBD)  algorithm to contract the
two-dimensional infinite lattice, with dimension of boundary matrix product
states (MPS) $D_{c}=3D$, and then measure the physical quantities following
iPEPS algorithm \cite{J. Jordan18}. In the simulation of spin-3/2
KHAF, we find $D_{c}=3D$ is sufficient to get the convergent and accurate
expectation value, however in the case of spin-3/2 XXZ model, $D_{c}=D^{2}$
is practically chosen.

Here we describe how to calculate thermodynamic properties with cluster update scheme.
In thermodynamical calculations, we use $\langle\hat{A}\rangle=Tr(\rho\hat{A})$ to get observations and therefore, we need to get the density operator $\rho$ at different temperature.

\begin{figure}
\includegraphics[width=0.9\linewidth,clip]{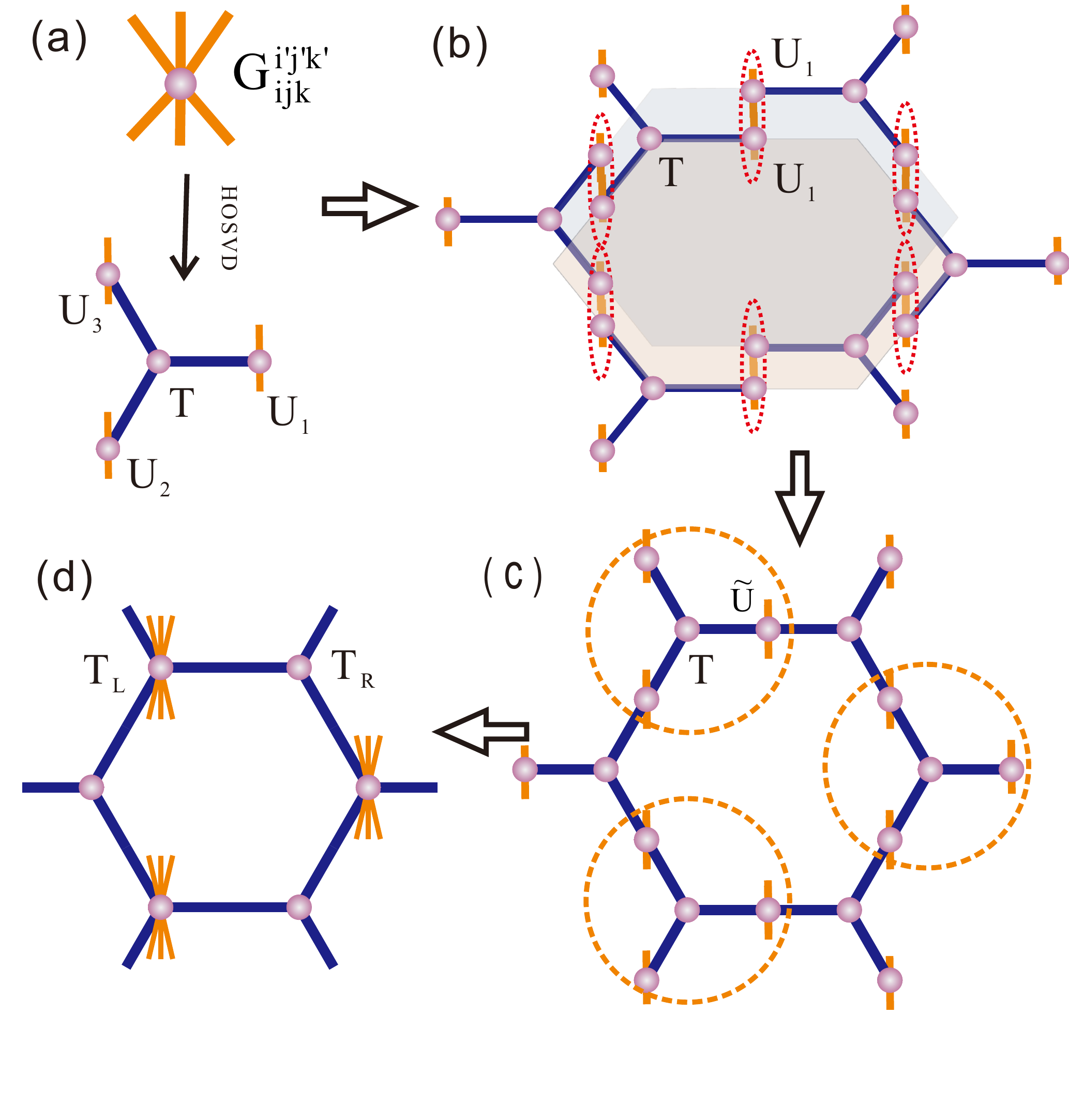}
\caption{(Color online) (a) Graphic illustration of a higher-order singular value decomposition. (b) Tensor network representation of the initial density operator. (c, d) After some transformations, one obtains the initial density operator.}
\label{cluster2}
\end{figure}

To get the initial density operator, we start from the Hamiltonian $H=\Sigma (H_{\triangle}+H_{\nabla})$, where $H_{\triangle,\nabla}$ are the local Hamiltonians of upper and down triangles, respectively.  The density operator can be expressed by $\rho=e^{-\beta H}\simeq[\Pi e^{-\tau H_{\triangle}}e^{-\tau H_{\nabla}}]^{K}$, where $\beta=\tau K$. Here we set $\tau=0.01$, and take $G_{\triangle,\nabla}=e^{-\tau H_{\triangle,\nabla}}$, where $G$ is a matrix with dimension $d^3\times d^3$ with $d$ the Hilbert space dimension of lattice spins. Then we make a higher-order singular value decomposition (HOSVD) on $G$ as shown in Fig. \ref{cluster2}(a):

\begin{equation}\label{4}
  (G)^{i',j',k'}_{i,j,k} = \sum_{x,y,z}T_{x,y,z}(U_{1})^{x}_{i,i'}(U_{2})^{y}_{j,j'}(U_{3})^{z}_{k,k'}.
\end{equation}

Then, the initial density operator $\rho_{T=1/\tau}=\Pi e^{-\tau H_{\triangle}}e^{-\tau H_{\nabla}}$ can be expressed as Fig. \ref{cluster2}(b). Note that the initial density operator $\rho_{T=1/\tau}$ is a two-layer tensor network which can merge together by contracting the shared indices between two conjoint tensors $U$ [see Eq. (\ref{5})]. One can get Fig. \ref{cluster2}(c) from Fig. \ref{cluster2}(b). In the last step, we can contract the tensors within the circles in Fig. \ref{cluster2} and obtain the tensor $T_{L}$. Figure. \ref{cluster2}(d) is the density operator $\rho_{T=1/\tau}$.

\begin{equation}\label{5}
  (\tilde{U})^{i,i''}_{x,x'} = \sum_{i'}U^{x}_{i,i'}U^{x'}_{i',i''}.
\end{equation}

It is seen that the tensor network representation of density operator is similar to that of ground-state wave function. What we have done in the ground state calculations can also apply to thermodynamic calculations. The initial temperature is chosen as $T/J=1/\tau$. We lower the temperature gradually by projecting $e^{-\tau H_{\triangle,\nabla}}$, and truncate the tensor network with cluster update scheme (as explained in ground state calculations) after every step of projection. At each temperature, we can calculate the observations by $\langle\hat{A}_{2T}\rangle=Tr(\rho_{T}\hat{A}\rho_{T})$. Considering the computational cost, here we just use local approximation to obtain $\langle\hat{A}_{2T}\rangle$ rather than global contraction, namely, we use six $\lambda's$ around the cluster tensors $N_{1}$, $N_{2}$ and $N_{3}$ to approximately represent the environment, as shown in Fig.\ref{cluster}(a).

\subsection{Full update}
\begin{figure}
\includegraphics[width=0.85\linewidth,clip]{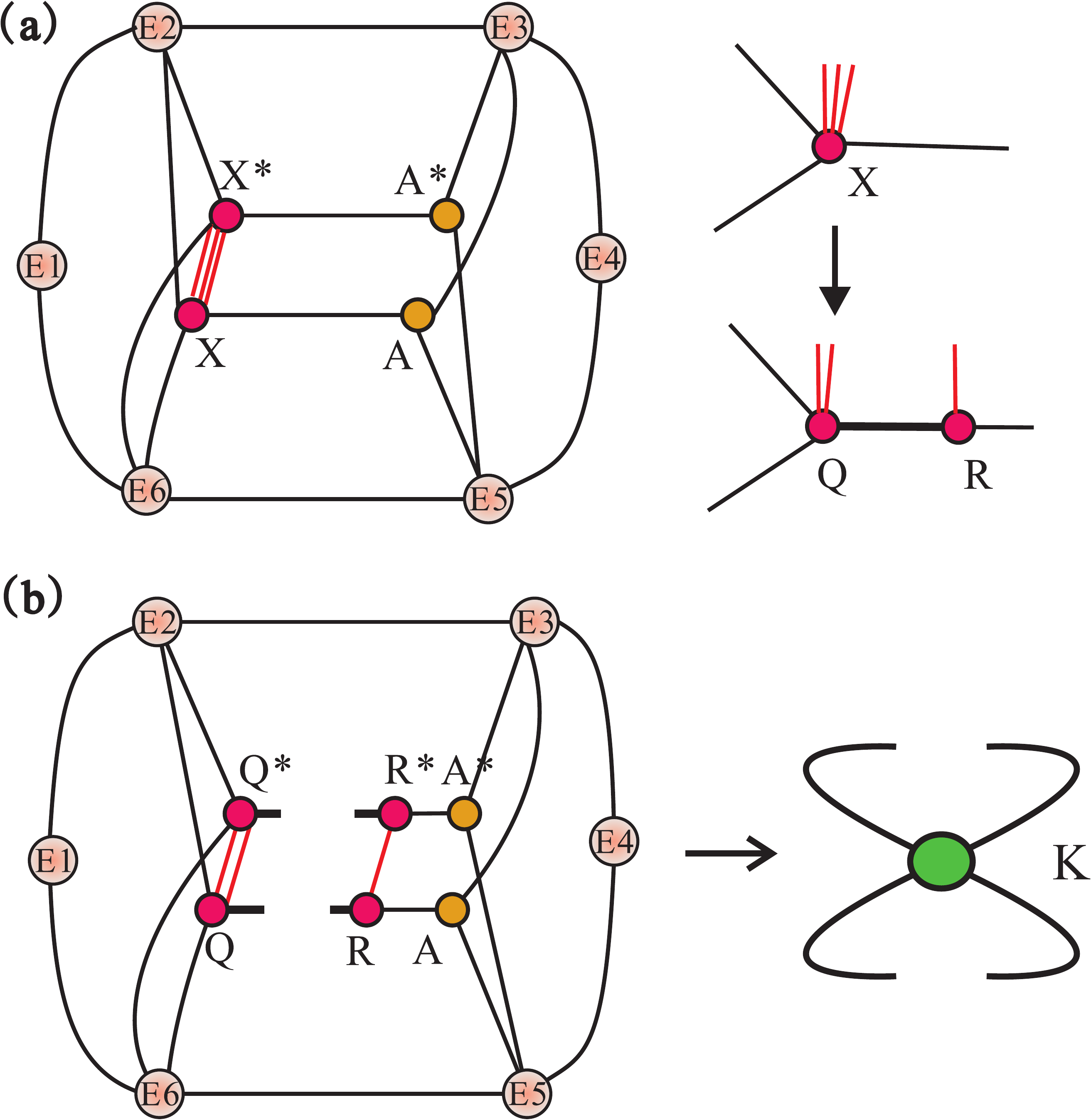}
\caption{(Color online) Graphic representation for the full update scheme. (a)
Tensors E1, E2, $\cdots$, E6 are environmental matrices obtained by the iTEBD
algorithm. Tensors X and A are the ``system" tensors that need to be transformed
and optimized, where tensor X has three physical indices but tensor A does not.
The right panel shows the QR decomposition of tensor X. (b) By cutting the bond
between tensors Q and R, and contracting the whole tensor cluster, one get an
environment tensor K, with which one optimizes the decimation matrix
variationally.}
\label{full}
\end{figure}

Different from the simple and cluster update schemes, full update scheme
contracts the whole two-dimensional tensor network to obtain the
global environment, rather than making ``local" approximation (in another word, a Bethe
lattice approximation \cite{W. Li32}) of the environment. However, such a
procedure increases the computing cost significantly. There are two popular
ways, the iTEBD and corner transfer matrix (CTM) methods, which are widely used
to contract the environment globally. Here we adopt the iTEBD method in the full
update calculations. We take the bond dimension of MPS as $\chi=3D$, which is
sufficient for the spin-3/2 KHAF. After contracting the environment, we get the
tensor structure shown in Fig. \ref{full} (a), where the tensors X and A
comprise the system and the tensors E1 to E6 represent the environment that we
obtained by iTEBD contractions. Next, we do a QR decomposition on tensor X, and
separate one physical index [right column in Fig. \ref{full} (a)]. Fig.
\ref{full} (b) gives an example along the direction of tensors X and A, which
leads to an increase of the bond dimension to $d \times D$ (for a spin-3/2
system, $d=4$). So we need to reduce the bond dimension from $d\times D$ to $D$,
to make the update procedure sustainable.

Then we optimize the truncation matrices variationally. By disconnecting the
bond between tensors Q and R, as shown in Fig. \ref{full} (b), and contracting
all environment tensors, we get a tensor K, which has four geometrical bonds
with dimension of $d\times D$. The subsequent process
for truncating the enlarged bond dimension follows Refs. [\onlinecite{J. Jordan18,H. N. Phien35,M. Lubasch36}], and the only difference is that we use
the bond-based truncation optimization, instead of site-based ones. In the
latter case, X and A tensors are to be optimized, while in the present scheme,
we find variationally the bond truncation matrix, starting with two identity
matrices, to decimate the enlarged geometric bonds. After the variational
optimization, we get a pair of truncation matrices $U_{L}$ and $U_{R}$ between
tensors Q and R. We can then proceed to find the truncation matrices along other
two directions of tensor X. Likewise, the tensors Y and Z can be truncated in a
similar way. When nine pairs of truncation matrices are at hand, we can perform
a truncation transformation on every geometric bond of the tensor network.
Generally, for each step of the imaginary-time projection and truncation, we
need to make only one full contraction procedure for the environment, but we have to
solve nine times the truncation matrix by the variational optimization.

\end{appendix}


\begin{thebibliography}{99}

\bibitem{V. Elser1}              V. Elser, Phys. Rev. Lett. \textbf{62}, 2405 (1989).
\bibitem{J. B. Marston2}         J. B. Marston and C. Zeng, J. Appl. Phys. \textbf{69}, 5 962 (1991).
\bibitem{S. Sachdev3}            S. Sachdev, Phys. Rev. B \textbf{45}, 12377 (1992).
\bibitem{S. Yan4}                S. Yan, D. A. Huse, and S. R. White, Science \textbf{332}, 1173 (2011).
\bibitem{S. Depenbrock5}         S. Depenbrock, I. P. McCulloch, and U. Schollw${\ddot{o}}$ck, Phys. Rev. Lett. \textbf{109}, 067201 (2012 ).
\bibitem{T.H.Han6}               T.-H. Han, J. S. Helton, S. Chu, D. G. Nocera, J. A. Rodriguez-Rivera, C. Broholm, Y. S. Lee, Nature \textbf{492}, 406 (2012).
\bibitem{Y.Iqbal7}               Y. Iqbal, F. Becca, S. Sorella, and D. Poilblanc, Phys. Rev. B \textbf{87}, 060405(R) (2013).
\bibitem{Z. Y. Xie8}             Z. Y. Xie, J. Chen, J. F. Yu, X. Kong, B. Normand, and T. Xiang, Phys. Rev. X \textbf{4}, 011025 (2014).
\bibitem{I. Rousochatzakis9}     I. Rousochatzakis, Y. Wan, O. Tchernyshyov, and F. Mila, Phys. Rev. B \textbf{90}, 100406(R) (2014).
\bibitem{T.Liu10}                T. Liu, S.-J. Ran, W. Li, X. Yan, Y. Zhao, and G. Su, Phys. Rev. B \textbf{89}, 054426 (2014).
\bibitem{W. Zhu11}               W. Zhu, S. S. Gong, and D. N. Sheng, Phys. Rev. B \textbf{92}, 014424 (2015).
\bibitem{T.H.Han12}              M. Fu, T. Imai, T.-H. Han, Y. S. Lee, Science \textbf{350}, 655 (2015).
\bibitem{T. Liu13}               T. Liu, W. Li, A. Weichselbaum, J. von Delft, and G. Su, Phys. Rev. B \textbf{91}, 060403(R) (2015).
\bibitem{W. Li14}                W. Li, A. Weichselbaum, J. von Delft, and H.-H. Tu, Phys. Rev. B \textbf{91}, 224414 (2015).
\bibitem{H. J. Changlani15}      H. J. Changlani and A. M. L¡§auchli, Phys. Rev. B \textbf{91}, 100407 (2015).
\bibitem{T. Picot16}             T. Picot and D. Poilblanc, Phys. Rev. B \textbf{91} , 064415 (2015).
\bibitem{W. Li17}                W. Li, S. Yang, M. Cheng, Z. X. Liu, and H. H. Tu, Phys. Rev. B \textbf{89}, 174411 (2014).
\bibitem{D. Ixert18}             D. Ixert, T. Tischler, and K. P. Schmidt, Phys. Rev. B \textbf{92}, 174422 (2015).

\bibitem{D. Ixert19}            T. Picot, M. Ziegler, R. Or\'{u}s, and D. Poilblanc, Phys. Rev. B \textbf{93}, 060407(R) (2016).

\bibitem{F. Verstraete29}        F. Verstraete and J. I. Cirac, arXiv:cond-mat/0407066; V. Murg, F. Verstraete, and J. I.
 Cirac, Phys. Rev. A \textbf{75}, 033605 (2007).
\bibitem{J. Jordan18}            J. Jordan, R. Or\'{u}s, G. Vidal, F. Verstraete, and J. I. Cirac, Phys. Rev. Lett. 101,
250602 (2008); R. Or\'{u}s and G. Vidal, Phys. Rev. B 80, 094403 (2009);

\bibitem{O. Gotze19}             O. G\"{o}tze, D. J. J. Farnell, R. F. Bishop, P. H . Y. Li, and J. Richter, Phys. Rev. B \textbf{84}, 224428 (2011).
\bibitem{A. L. Chernyshev25}     A. L. Chernyshev and M. E. Zhitomirsky, Phys. Rev. Lett. \textbf{113}, 237202 (2014).
\bibitem{A. L. Chernyshev20}     A. L. Chernyshev and M. E. Zhitomirsky, Phys. Rev. B \textbf{92}, 144415 (2015).
\bibitem{C. Mondelli21}          X. Obradors, A. Labarta, A. Isalgue, J. Tejada, J. Rodriguez, and M.
                                 Pernet, Solid State Commun. \textbf{65}, 189 (1988); A. P. Ramirez, G. P. Espinosa, and A. S. Cooper, Phys.
                                 Rev. Lett. \textbf{64}, 2070 (1990); A. P. Ramirez, G. P. Espinosa, and A. S. Cooper, Phys. Rev. B \textbf{45}, 2505 (1992); C.
                                 Broholm, G. Aeppli, G. P. Espinosa, and A. S. Cooper, Phys. Rev. Lett. \textbf{65}, 3173 (1990); Y. J.
                                 Uemura, A. Keren, K. Kojima, L. P. Le, G. M. Luke, W. D. Wu, Y. Ajiro, T. Asano, Y. Kuriyama, M.
                                 Mekata, H. Kikuchi, and K. Kakurai, Phys. Rev. Lett. \textbf{73}, 3306 (1994); C. Mondelli, H. Mutka, C.
                                 Payen, B. Frick, and K. H. Andersen, Physica B \textbf{284}, 1371-1372 (2000).
\bibitem{S. E. Dutton22}         S. E. Dutton, E. D. Hanson, C. L. Broholm, J. S. Slusky, and R. J. Cava,
                                     Journal of Physics-Condensed Matter \textbf{23}, 386001 (2011).
\bibitem{D. Bono23}              I. S. Hagemann, Q. Huang, X. P. A. Gao, A. P. Ramirez, and R. J.
                                 Cava, Phys. Rev. Lett. \textbf{86}, 894 (2001); D. Bono, P. Mendels, G. Collin, N. Blanchard, F. Bert, A.
                                   Amato, C. Baines, and A. D. Hillier, Phys. Rev. Lett. \textbf{93}, 187201 (2004).
\bibitem{Cr-jarosite24}          M. G. Townsend, G. Longworth, and E. Roudaut, Phys. Rev. B \textbf{33}, 4919 (1986); A. Keren, K. Kojima, L. P. Le, G. M. Luke, W. D. Wu, Y. J. Uemura, M. Takano, H. Dabkowska, and M. J. P. Gingras, Phys. Rev. B \textbf{53}, 6451 (1996); T. Inami, T. Morimoto, M. Nishiyama, S. Maegawa, Y. Oka, and H. Okumura, Phys. Rev. B \textbf{64}, 054421 (2001); K. Okuta, S. Hara, H. Sato, Y. Narumi, K. Kindo,  J. Phys. Soc. Jpn. \textbf{80}, 063703(2011).
\bibitem{lee}                    S.-H. Lee, C. Broholm, M. F. Collins, L. Heller, A. P. Ramirez, Ch. Kloc, E. Bucher, R. W. Erwin, and N. Lacevic,  Phys. Rev. B \textbf{56}, 8091 (1997).
\bibitem{A. B. Harris}           A. B. Harris, C. Kallin, and A. J. Berlinsky, Phys. Rev. B 45, 2899 (1992).
\bibitem{A. Chubukov}            A. Chubukov, Phys. Rev. Lett. 69, 832 (1992).
\bibitem{O. Gotze26}             O. G\"{o}tze and J. Richter, Phys. Rev. B \textbf{91}, 104402 (2015).
\bibitem{J. Oitmaa27}            J. Oitmaa and R. R. P. Singh, Phys. Rev. B \textbf{93}, 014424 (2016).
\bibitem{A. Lohmann28}           A. Lohmann, H. J. Schmidt, and J. Richter, Phys. Rev. B \textbf{89}, 014415 (2014).

\bibitem{Masuo Suzuki30}         M. Suzuki, Prog. Theor. Phys. \textbf{56}, 1454 (1976).
\bibitem{H. C. Jiang31}          H. C. Jiang, Z. Y. Weng, and T. Xiang, Phys. Rev. Lett. \textbf{101}, 090603 (2008).
\bibitem{W. Li32}                W. Li, J. von Delft, and Tao Xiang, Phys. Rev. B \textbf{86}, 195137 (2012).
\bibitem{L. Wang33}              L. Wang and F. Verstraete, arXiv:1110.4362 (2011).
\bibitem{H. N. Phien35}          H. N. Phien, J. A. Bengua, H. D. Tuan, P. Corboz, and R. Or\'{u}s, Phys. Rev. B \textbf{92}, 035142 (2015).
\bibitem{M. Lubasch36}           M. Lubasch, J. I. Cirac, and M. C. Banuls, Phys. Rev. B \textbf{90}, 064425 (2014).
\bibitem{ran38}                  S.-J. Ran, W. Li, B. Xi, Z. Zhang, and G. Su, Phys. Rev. B \textbf{86}, 134429 (2012).
\bibitem{R. Orus32}              R. Or\'us and G. Vidal, Phys. Rev. B \textbf{78}, 155117 (2008).
\bibitem{schnack}                J. Schnack, H.-J. Schmidt, J. Richter, and J. Schulenburg, Eur. Phys. J. B \textbf{24}, 475 (2001); J. Schulenburg, A. Honecker, J. Schnack, J. Richter, and H.-J. Schmidt, Phys. Rev. Lett. \textbf{88}, 167207 (2002).
\bibitem{Mermin37}               N. D. Mermin and H. Wagner, Phys. Rev. Lett. \textbf{17}, 1133 (1966).
\bibitem{Li-2011}                W. Li, S.-J. Ran, S.-S. Gong, Y. Zhao, B. Xi, F. Ye, and G. Su, Phys. Rev. Lett. \textbf{106}, 127202 (2011).
\bibitem{Ran-2013}               S.-J. Ran, B. Xi, T. Liu, and G. Su, Phys. Rev. B \textbf{88}, 064407 (2013).

\end{thebibliography}
\end{document}